# A Morphological Adaptation Approach to Path Planning Inspired by Slime Mould

Jeff Jones*

*Centre for Unconventional Computing, University of the West of England, Coldharbour Lane, Brisol, BS16 1QY, UK.*

Path planning is a classic problem in computer science and robotics which has recently been implemented in unconventional computing substrates such as chemical reaction-diffusion computers. These novel computing schemes utilise the parallel spatial propagation of information and often use a two-stage method involving diffusive propagation to discover all paths and a second stage to highlight or visualise the path between two particular points in the arena. The true slime mould *Physarum polycephalum* is known to construct efficient transport networks between nutrients in its environment. These networks are continuously remodelled as the organism adapts its body plan to changing spatial stimuli. It can be guided towards attractant stimuli (nutrients, warm regions) and it avoids locations containing hazardous stimuli (light irradiation, repellents, or regions occupied by predatory threats). Using a particle model of slime mould we demonstrate scoping experiments which explore how path planning may be performed by morphological adaptation. We initially demonstrate simple path planning by a shrinking blob of virtual plasmodium between two attractant sources within a polygonal arena. We examine the case where multiple paths are required and the subsequent selection of a single path from multiple options. Collision-free paths are implemented via repulsion from the borders of the arena. Finally, obstacle avoidance is implemented by repulsion from obstacles as they are uncovered by the shrinking blob. These examples show proof-of-concept results of path planning by morphological adaptation which complement existing research on path planning in novel computing substrates.

**Keywords:** path planning; morphological computation; slime mould; collective computation; collision avoidance; robotics

## 1. Introduction - Path Planning: Unconventional Computing Approaches

Path planning (or motion planning) is a common application of computer science and robotics where a path has to be found between points (typically two points, source and destination point) within an arena. The representation of the arena may already be known or may be discovered by localisation and mapping methods (in this paper we consider examples where the arena layout is known in advance). The resultant path should be short, minimising distance between the points. Other constraints may also apply, such as requiring paths of sufficient width, avoiding walls, avoiding obstacles, or minimising the number of turns.

Unconventional computing seeks to utilise the computing potential of natural physical systems to solve useful problems. Since these systems are localised in space, they typically use different mechanisms to classical approaches. In recent years physical propagation through space in chemical substrates has been used as a search strategy. Babloyantz first suggested that travelling wave-fronts from chemical reactions in excitable media could be used to approximate spatial problems (Babloyantz and Sepulchre 1991). Wave propagation in the Belousov-Zhabotinsky (BZ) chemical reaction was subsequently used to discover the path through a maze (Steinbock, Tóth, and Showalter 1995). In this research a trigger wave was initiated at the bottom left corner of a maze and its propagating wave front recorded by time-lapse photography. Direction of wave propagation was calculated from the collective time-lapse information to give vectors which indicated the direction of the travelling wave. The path from any point on the maze to the exit (the source of the diffusion) was followed by tracking backwards (using the vector information) to the source.

---

*Email: jeff.jones@uwe.ac.uk

Wave-front propagation generates a solution from any (and indeed *every*) point in the arena. Branching paths (for example around obstacles) are searched in parallel and the solution time is dependent on the spatial size (in terms of maximum path length) of the arena and the wave-front propagation speed. Although computationally efficient, a direct spatial encoding of the problem (arena, desired start and end points) must be stored, as opposed to a more compact graph or grid encoding in classical approaches.

Reading the output of the parallel calculations is not a simple approach using chemical substrates. Although the propagating wave solves the shortest path for all points in the arena, finding and tracking the desired path from start to end point requires separate processes. Different approaches have been attempted including image processing (Rambidi 2005), using two wave-fronts in both directions (Agladze et al. 1997), and hybrid chemical and cellular automata approaches (Adamatzky and de Lacy Costello 2003). More recently, a direct visual solution to path planning was devised in which an oil droplet (exploiting convection currents and surface tension effects) migrated along a pH gradient formed within a maze to track the shortest path through the maze (Lagzi et al. 2010).

In this paper we continue the exploration of material computation by morphological adaptation seen in (Jones and Adamatzky 2014a) and (Jones and Adamatzky 2014b) and examine its application to path planning. Taking inspiration from the behaviour of slime mould, we use a large sheet, or 'blob' of virtual slime mould which is located within an arena in which a path between two points (represented by attractants) must be found. By shrinking this blob over time, it withdraws from the confines of the arena boundary and adapts its shape to connect the start and end points of the path. We give an overview of the behaviour of slime mould as an inspiration for this method in Section 2. An overview of the agent-based model underlying the approach is given in Section 3. Examples of the shrinkage method are given in Section 4, along with more challenging additions to the problem such as multiple-path options, collision-free paths and obstacle avoidance. We conclude in Section 5 by summarising the approach and its contribution to unconventional computing methods of path planning in terms of its simplicity.

## 2. Computing by Morphological Adaptation in Slime Mould

The giant single-celled amoeboid slime mould *Physarum polycephalum*, has been the subject of intense research into unconventional computing substrates due to its relative biological simplicity and complex behaviour. In the plasmodium stage of its complex life cycle the organism forages towards, engulfs and consumes micro-organisms growing on vegetative matter. When presented with a distributed spatial configuration of nutrients (for example oat flakes) the plasmodium forms a network of protoplasmic tubes connecting the nutrients. This is achieved without recourse to any specialised neural tissue. The organism dynamically adapts its morphology to form efficient paths (in terms of a trade-off between overall distance and resilience to random damage) between the food sources (Nakagaki et al. 2004; Nakagaki and Guy 2007; Nakagaki et al. 2007).

Research into computation by *Physarum* was initiated by Nakagaki, Yamada and Toth, who reported the ability of the *Physarum* plasmodium to solve a simple maze problem (Nakagaki, Yamada, and Toth 2000). It has since been demonstrated that the plasmodium successfully approximates spatial representations of various graph problems (Nakagaki et al. 2004; Shirakawa et al. 2009; Adamatzky 2008; Jones 2011a), combinatorial optimisation problems (Aono and Hara 2007; Jones 2011b; Jones and Adamatzky 2014a), construction of logic gates and adding circuits (Tsuda, Aono, and Gunji 2004; Jones and Adamatzky 2010; Adamatzky 2010), and spatially represented logical machines (Adamatzky 2007; Adamatzky and Jones 2010).

In its plasmodium state *Physarum* does not approximate the internal area of a shape whose borders are defined by the placement of nutrients. This is because the plasmodium spontaneously forms networks spanning the sources. If a plasmodium is inoculated as a solid sheet of material, the sheet is soon transformed into a network structure by competitive flux of material within the sheet (Nakagaki et al. 2004). Furthermore, the shape adaptation of the plasmodium is restricted by adhesion to the slime capsule generated by the plasmodium during its growth and adaptation. It is thus physically impractical to force a freely foraging plasmodium to conform to a solid shape.

## 3.  Morphological Adaptation in a Model of Slime Mould

Nevertheless, the material computation embodied within *Physarum* presents intriguing possibilities towards generating novel spatially represented methods of unconventional computation. We have previously explored *Physarum*-inspired mechanisms of material computation using a multi-agent particle model of the *Physarum* plasmodium which performs computation by morphological adaptation. Most recently we have used this approach to approximate the Travelling Salesman Problem (TSP) (Jones and Adamatzky 2014a) and to perform data smoothing and generate spline curves (Jones and Adamatzky 2014b). In these approaches we used a large population, or 'blob' of a virtual plasmodium material to perform the spatial computation. The morphological adaptation of the blob over time was constrained by the presence of attractant sources representing problem data points. The final output of the model was contained in the shape of the network (the peripheral path of the blob for the TSP, and the course of the blob path for spline curves and data smoothing problems). As a brief overview of the model (a more detailed description is given in the appendix) the material is composed of thousands of simple mobile multi-agent particles interacting together within a 2D diffusive lattice. Each particle senses the concentration of a generic 'chemo-attractant' substance diffusing within the lattice and each agent also deposits the same substance within the lattice upon successful forward movement. The multi-agent population collectively exhibits emergent properties of cohesion and shape minimisation as a results of the low-level particle interactions. The pattern formation and network adaptation properties of small populations of the material were discussed in (Jones 2010) and were found to reproduce a wide range of Turing-type reaction-diffusion patterning.

## 4.  Results: Path Planning by Collective Morphological Adaptation

### 4.1.  *Simple Path Planning*

We placed a large population of particles within the confines of a 2D arena (Fig. 1a), so that the virtual plasmodium completely filled the arena (Fig. 1b). Start and end points of the path were represented by projection of attractant into the arena at their respective locations. The virtual plasmodium was attracted to these start and end points. The population size was reduced by adjusting the parameters governing the growth and shrinkage in favour of shrinkage. The collective 'blob' began to shrink and, as it did so, adapted its shape to maintain connectivity to the start and end points and conform to to the borders of the arena (Fig. 1c-e). Any extraneous pseudopodium-like appendages were withdrawn until only a single path connected the start and end points, forming the shortest path between the two points (Fig. 1f). This path was composed of a thin band of particles.

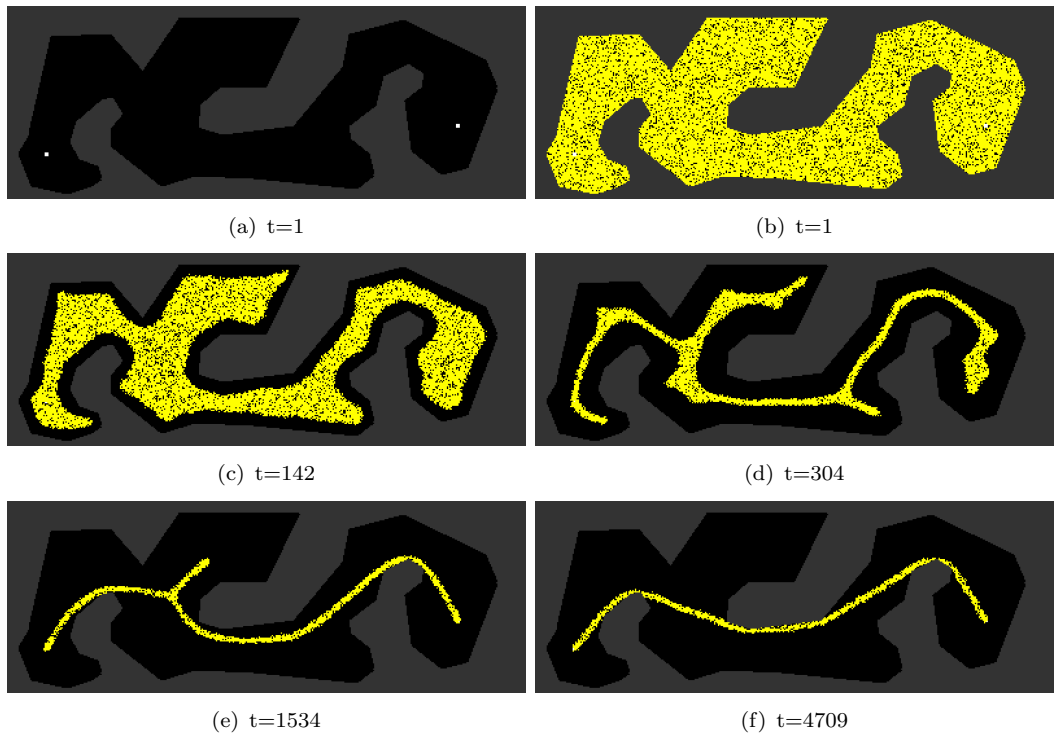

(a) t=1     (b) t=1

(c) t=142     (d) t=304

(e) t=1534     (f) t=4709

Figure 1.  Approximation of shortest path between two points by morphological adaptation in a virtual slime mould. a) 2D Arena defined by borders (grey), habitable regions (black) and start and end points (white), b) initialisation of virtual plasmodium in habitable region, c-f) shrinkage of blob causes adaptation of shape and attraction to points, ultimately forming the shortest path between the points.

### 4.2.  *Multiple Path Planning and Many-to-one Path Selection*

Most applications for path planning involve finding a path between two points. However it may be necessary to calculate a connecting path between multiple points. This can be achieved in the morphological adaptation approach by having multiple nutrient attractant sources. When a large blob is placed within a polygon arena containing multiple sources, it retains its connection to all points as it shrinks (Fig. 2). The final network path (Fig. 2d) indirectly connects all source points. Note that the connections between the sources are not simple edges to and from each point. This is not possible because straight edges would pass through the protruding arena boundaries. Instead there is a core curved path running between the outermost points (1 and 4) with pseudopodium-like extensions protruding from this path to connect the inner points (2 and 3). Note that this core path passing between four points differs from that of Fig. 1f which passes between only two outer points. The connection between the core path and the two inner points (2 and 3) distorts the core path in the direction of the inner points.

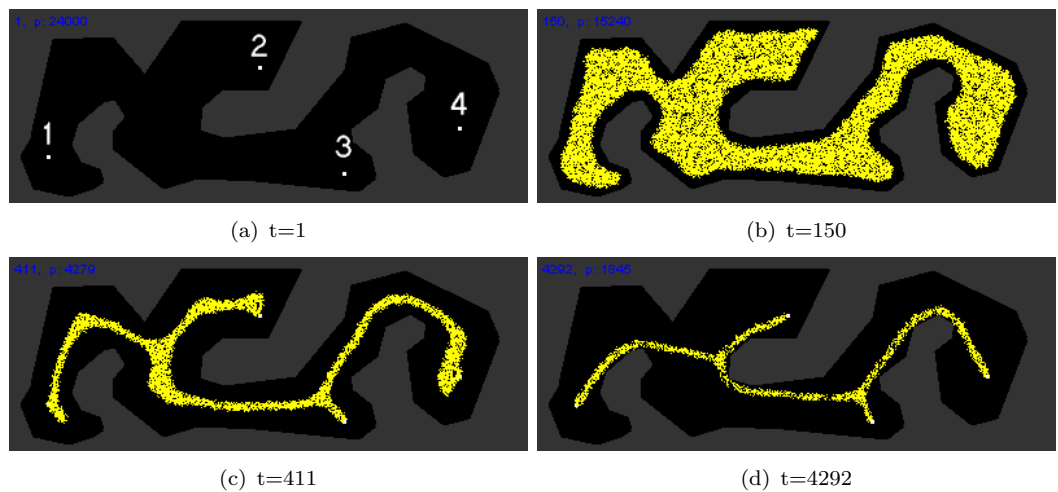

(a) t=1    (b) t=150

(c) t=411    (d) t=4292

Figure 2. Multiple source path planning by morphological adaptation. a) arena with four locations (numbered white points), b-d) shrinkage of virtual plasmodium yields a network with a core path extending from the outermost points with extensions connecting the inner points.

Given this path connecting all four points, how does the morphological adaptation method respond to the removal of certain path options? In Figs. 3 and 4 we demonstrate the effect of removing different path options. Beginning with the path connecting four points (Fig. 3a) we remove points 2 and 3 by deleting the attractant sources at these locations from the lattice. The pseudopodium-like projections connecting these branches to the core path both retract into the core path due to the lack of attractant from these sources (Fig. 3b,c). The retracting projections merge with the main flow of particles in the core path. When retraction of these branches is complete the core path connecting points 1 and 4 continues its adaptation to adopt a minimal path between the outer points following the contours of the arena boundary (Fig. 3d).

If we remove different source points from the same starting configuration, the adaptation takes a different course. Fig. 4 shows the results of a different experiment with the same four points initially connected (Fig. 4a). When source points 2 and 4 are removed from the lattice the pseudopodium withdrawal again commences (Fig. 4b,c) and the final remaining path adopts a minimal connection between points 1 and 3 (Fig. 4d).

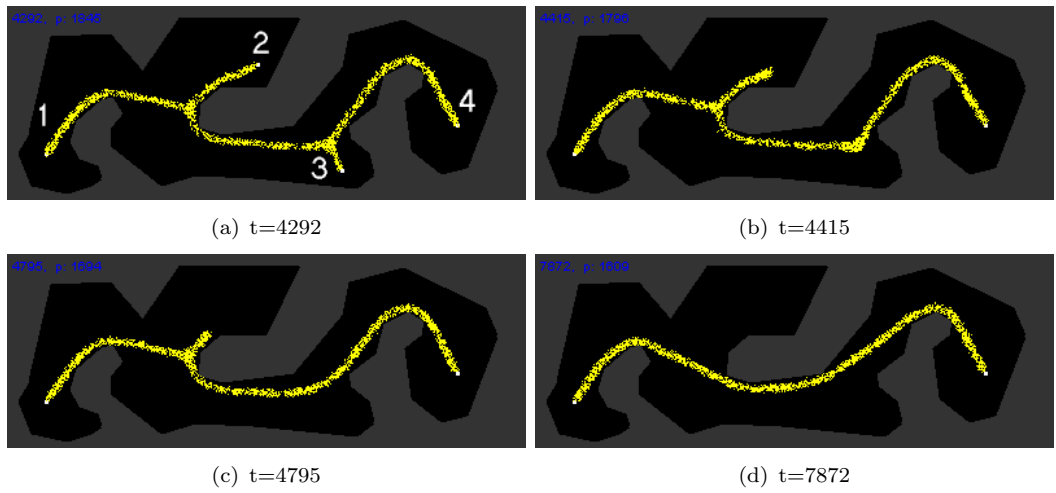

Figure 3. From multiple paths to single path by morphological adaptation. a) multiple paths connecting four attractants (numbered), b-d) removal of attractant source from nodes 2 and 3 causes withdrawal of pseudopodia from previous sources and ultimately a single path formed between nodes 1 and 4.

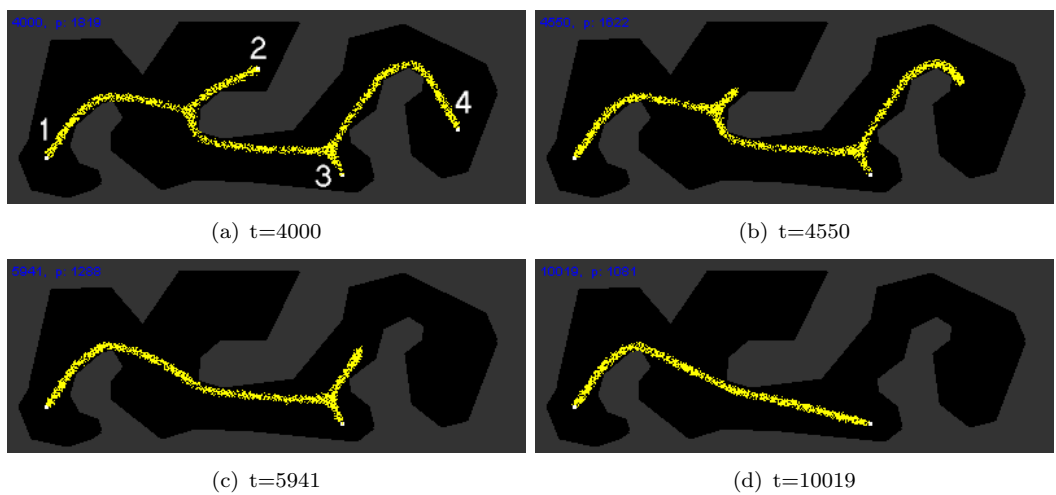

Figure 4. From multiple paths to single path by morphological adaptation. a) multiple paths connecting four attractants (numbered), b-d) removal of attractant source from nodes 2 and 4 causes withdrawal of pseudopodia from previous sources and ultimately a single path formed between nodes 1 and 3.

### 4.3. *Collision-free Paths via Repulsion*

In many applications a collision-free path may be required, for example if the desired path has to avoid close proximity to walls. To achieve this method by morphological adaptation we represented the walls of the arena as repellent sources (repellent sources project negatively weighted values into the diffusive lattice). We used the same arena as in earlier experiments, but with different start and end points (Fig. 5a,b). As the blob shrunk it formed the shortest path (following the walls) when repellent diffusion was not activated (Fig. 5c-f). When repellent diffusion from the arena walls was activated the virtual plasmodium still maintained its connectivity to the start and end points but also avoided the diffusing repellent values projecting from the walls of the arena (Fig. 5g). Further increasing the concentration of the repellent source increased the distance of the path from the walls (Fig. 5h).

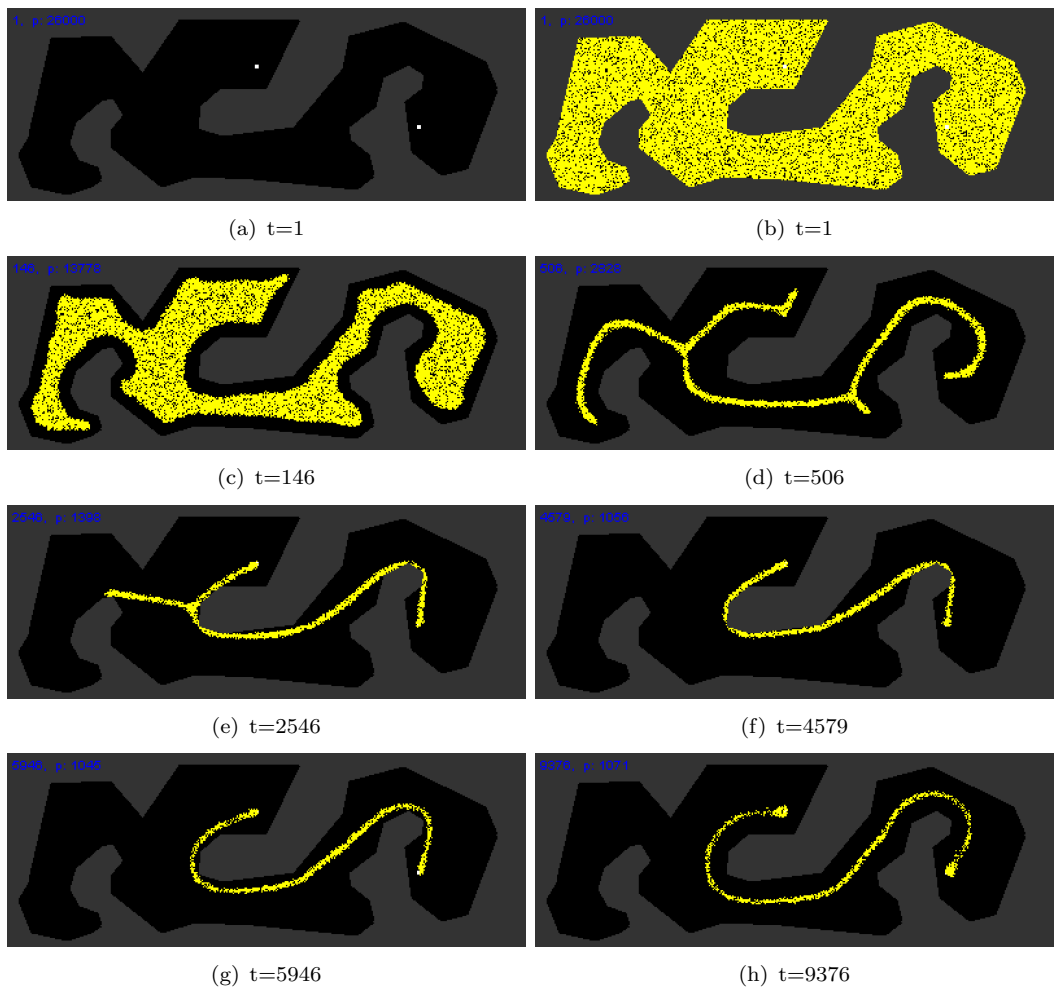

(a) t=1    (b) t=1

(c) t=146    (d) t=506

(e) t=2546    (f) t=4579

(g) t=5946    (h) t=9376

Figure 5. Approximation of collision-free shortest path by morphological adaptation and repulsion. a) 2D Arena defined by borders (grey), habitable regions (black) and start and end points (white), b) initialisation of virtual plasmodium in habitable region, c-f) shrinkage of blob causes adaptation of shape and attraction to points, forming the shortest path between the points, g) repulsion field emitted from arena walls causes virtual plasmodium to avoid wall regions, forming a collision-free path, h) increasing concentration of repulsion field causes further adaptation of the virtual plasmodium away from walls.

### 4.4. Obstacle Avoidance and Preventing Multiple Paths

The response of the morphological adaptation in the presence of obstacles is shown in Fig. 6. Again the model is initialised within the habitable region of the arena. When the virtual plasmodium shrinks and adapts its shape in the presence of obstacles we see that multiple paths are formed around the obstacles which connect the start and end points (Fig. 6f). Even if we projected repellents from these obstacles, the effect would only be to widen the distance of the multiple paths from these obstacles, and not to form a single path.

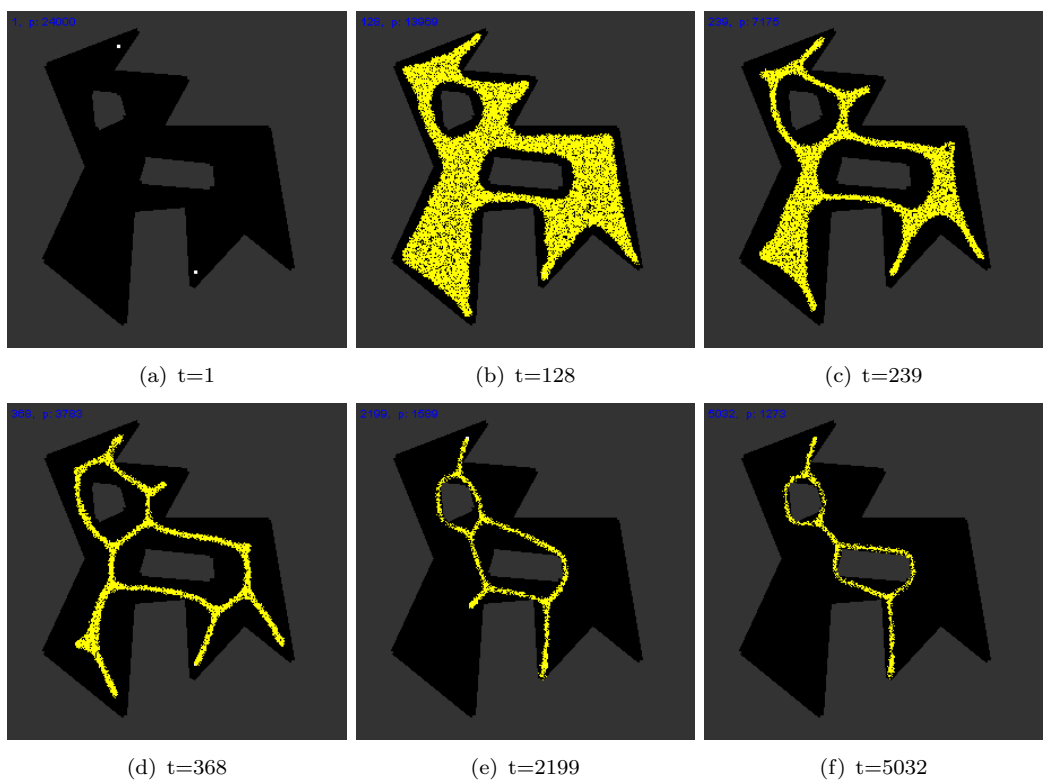

(a) t=1    (b) t=128    (c) t=239

(d) t=368    (e) t=2199    (f) t=5032

Figure 6. Addition of obstacles results in multiple paths connecting start and end points. a) 2D Arena defined by borders and obstacles (grey), habitable regions (black) and start and end points (white), b-e) shrinkage of virtual plasmodium within habitable region, f) shrinkage of virtual plasmodium around obstacles causes multiple paths around obstacles on the path between start and end points.

The final shrunken multiple paths connecting the nutrient sources are observed because they actually already existed upon initialisation, as the blob was initialised around the obstacles (see Fig. 6b). To ensure only a single path is generated we devised a two-part repulsion mechanism. The first part of the mechanism occurs by initialising the blob to cover the *entire* arena (including the obstacles). This part in isolation would not solve the problem of multiple paths, however: If the obstacles repelled the blob immediately then the virtual plasmodium would simply flee the obstacle regions from all directions and multiple paths would still be retained. The second part of the mechanism ensures that only a single path is retained. The shrinkage process is performed more slowly and we generate repellent fields only from obstacles (more specifically, exposed *fragments* of large obstacles) that have been partially uncovered by the shrinkage of the blob.

The mass of the blob is thus shifted away from obstacles by their emergent repulsion field. Because the blob shrinks slowly inwards from the outside of the arena obstacles are slowly uncovered and the repulsion field further pushes the blob inwards until a single path connecting the source attractants is formed. The shrinkage and repulsion mechanism is illustrated in Fig. 7 where the arena (including obstacles) is completely covered by a large mass of particles comprising the virtual plasmodium (Fig. 7b). The blob shrinks inwards as the periphery of the blob is drawn inwards (Fig. 7c). When an obstacle is partially uncovered repellent is projected into the diffusive lattice at exposed obstacle fragments (Fig. 7d, arrowed). The blob at these regions is repelled and moves away from the exposed obstacle fragment. The shrinkage process continues and when a larger obstacle is partially exposed the repellent projected into the lattice again causes the blob to move away from this region (Fig. 7f, arrowed). Further exposure of this large lower obstacle causes the blob to continue to be repelled away (Fig. 7g) until eventually only a single path remains which connects the source attractants and threads between the obstacles (Fig. 7h).

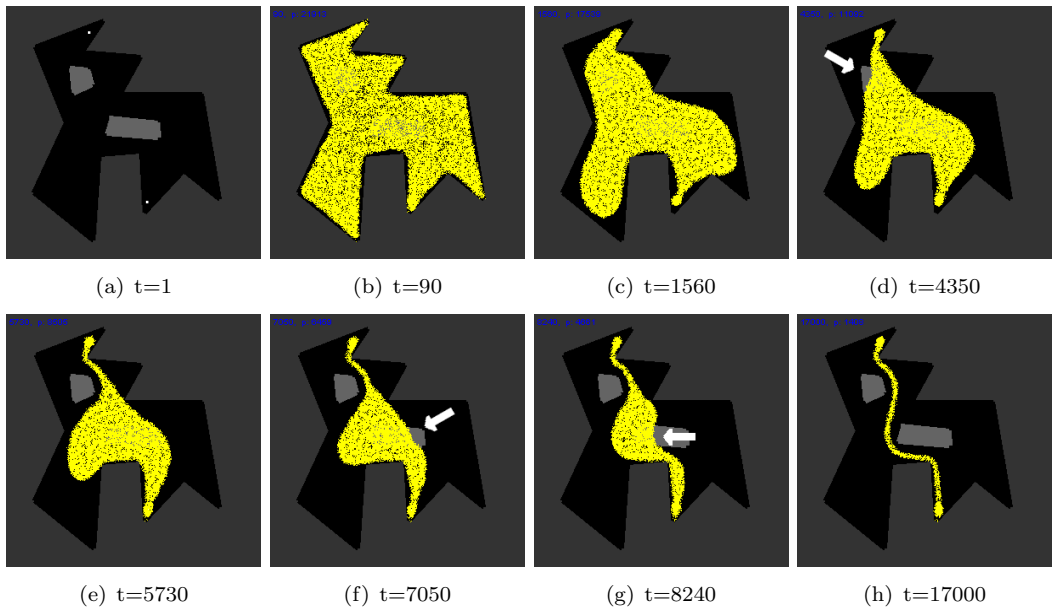

(a) t=1   (b) t=90   (c) t=1560   (d) t=4350
(e) t=5730   (f) t=7050   (g) t=8240   (h) t=17000

Figure 7. Mechanism of shrinkage combined with repulsion at exposed obstacle fragments generates a single path. a) arena with habitable areas (black), inhabitable areas (dark grey), obstacles light grey and path source locations (white). b) blob initialised on entire arena, including obstacles, c) gradual shrinkage of blob, d) exposure of obstacle fragment generates repellent field at exposed areas (arrow), e) blob moves away from repellent field of obstacle, f) lower obstacle is exposed causing repellent field at these locations (arrow), g) further exposure causes migration of blob away from these regions (arrow), h) final single path connects source points whilst avoiding obstacles.

In the presence of a large number of obstacles, the repulsion field emanating from newly-exposed obstacles acts to deform the shrinkage of the virtual plasmodium. The mass of particles is deformed both by the attractant stimuli from the start and end points of the path (Fig. 8d) and the gradual exposure of the obstacles as the blob shrinks. Fig. 8 shows the deformation of the blob and also the changing concentration gradient field as the shrinkage continues, until only a collision-free path between the obstacles remains.

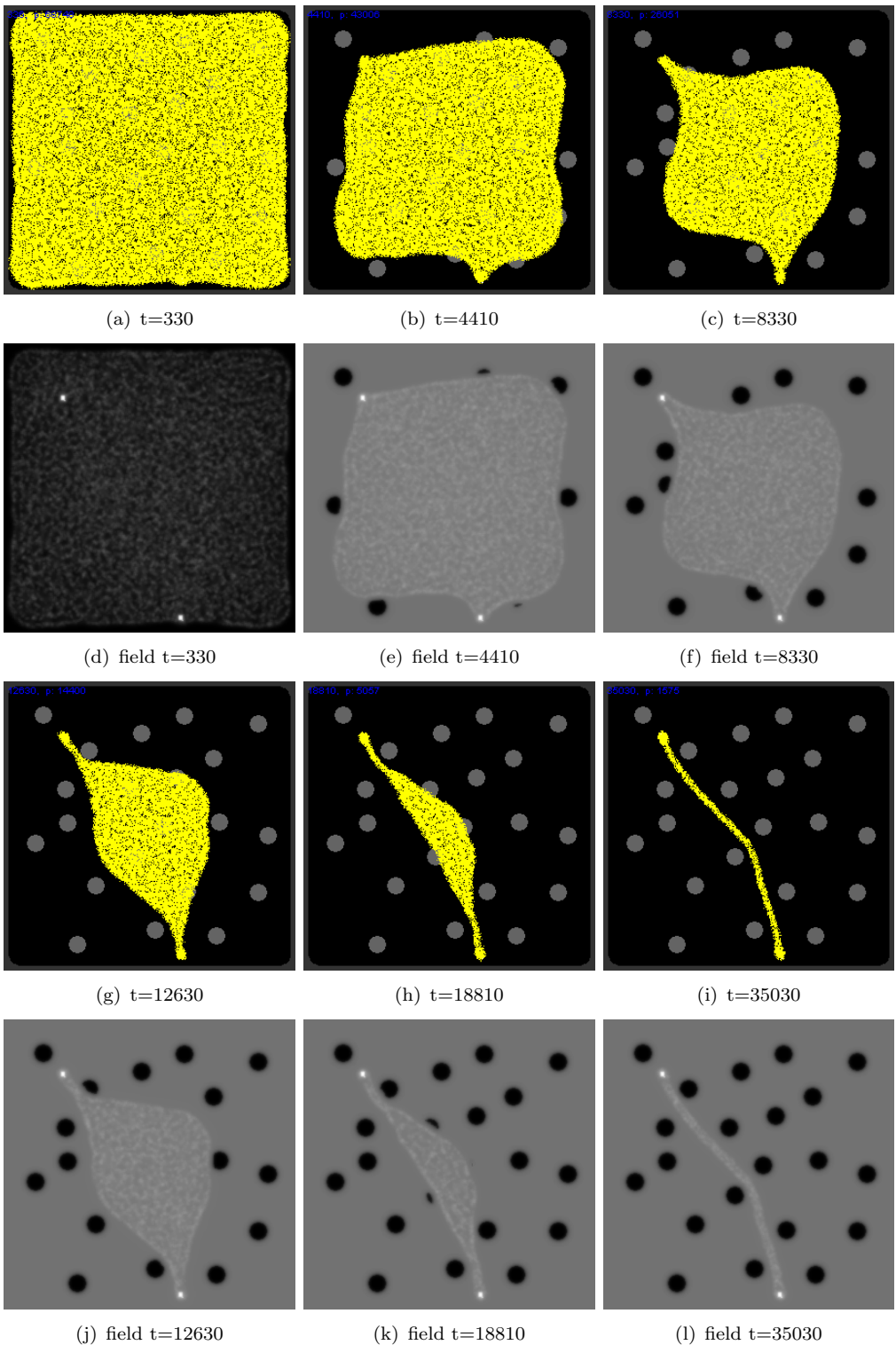

Figure 8. Shrinkage and exposed repulsion method in complex obstacle field. (a-c) and (g-i) uniform shrinkage of blob is distorted by attraction to start and end stimuli and repulsion from exposed obstacles, (d-f) and (j-l) visualisation of gradient field showing attractants at start and end stimuli (bright spots), blob (mid-grey mass) and repulsion field from exposed obstacles (dark circles). Greyscale gradient field images transformed by gamma correction ($\gamma = 0.6$) to improve clarity.

## 5. Conclusions

We have demonstrated how unconventional computation of path planning problems may be performed directly in 2D space by morphological adaptation in a virtual material inspired by the adaptation of slime mould *Physarum polycephalum*. Unlike previous implementations of path planning problems in chemical substrates the method does not rely on a two-stage computation (one stage to perform the computation, another stage to highlight the path). The method computes a simple path with only two attractant sources. Multiple paths were represented by having more than two attractant sources and a single path was selected between two of these sources by removal of redundant sources. Collision-free paths were discovered by the simultaneous addition of repellent sources at arena boundaries. Obstacle avoiding paths were discovered using a mechanism whereby obstacles were represented by a gradual exposure of repellent sources. The contribution of this method is in the simplicity of the approach: the behaviour of the shrinking blob is distributed within the material itself and emerges from the simple and local interactions between the particles which comprise the blob. The path finding process is governed, to a large extent, by the spatial configuration of the arena and the obstacles within the arena. Since the blob initially occupies all of the space within the arena the path finding method may be described as subtractive — all redundant or inefficient paths are removed during the shrinkage process. This is achieved by withdrawal of pseudopodia (for example from dead-ends in the arena) and also by displacement of the blob by the repellent fields emitted from the gradually exposed obstacles. Unlike chemical-based approaches the method is not initiated at either the path start or end points but is initialised by shrinkage from the arena boundary. Diffusion from the source points still occurs but is merely used to anchor the blob material at these points and does not require propagation of the diffusion front throughout the entire arena. Likewise, repellent diffusion occurs from the boundary (for collision-free paths) and from obstacles (for obstacle-avoiding paths) but this diffusion also is only local and does not require propagation throughout the entire arena.

## Funding

This paper was supported by the EU research project "Physarum Chip: Growing Computers from Slime Mould" (FP7 ICT Ref 316366).

## Supplemental material

Supplementary video recordings visualising the shrinkage and adaptation of the model plasmodium in path planning examples can be found at: http://uncomp.uwe.ac.uk/jeff/pathplanning.htm.

## 6.  Appendix. Particle Model Description

The multi-agent particle approach to generate the behaviour of the virtual material blob uses a population of indirectly coupled mobile particles with very simple be-haviours, residing within a 2D diffusive lattice which stores particle positions and the concentration of a generic diffusive factor referred to as chemo-attractant. Collective particle positions represent the global pattern of the blob and collective particle motion represents flux within the blob. The particles act independently and iteration of the particle population is performed randomly to avoid any artifacts from sequential ordering.

### 6.1. Generation of Emergent Blob Cohesion and Morphological Adaptation

The behaviour of the particles occurs in two distinct stages, the sensory stage and the motor stage. In the sensory stage, the particles sample their local environment using three forward biased sensors whose angle from the forwards position (the sensor angle parameter, SA), and distance (sensor offset, SO) may be parametrically adjusted (Fig. 9a). The offset sensors generate local coupling of sensory inputs and movement to generate the cohesion of the blob. The SO distance is measured in pixels and a minimum distance of 3 pixels is required for strong local coupling to occur. During the sensory stage each particle changes its orientation to rotate (via the parameter rotation angle, RA) towards the strongest local source of chemo-attractant (Fig. 9b). After the sensory stage, each particle executes the motor stage and attempts to move forwards in its current orientation (an angle from 0–360°) by a single pixel forwards. Each lattice site may only store a single particle and particles deposit chemo-attractant into the lattice only in the event of a successful forwards movement. If the next chosen site is already occupied by another particle the move is abandoned and the particle selects a new randomly chosen direction.

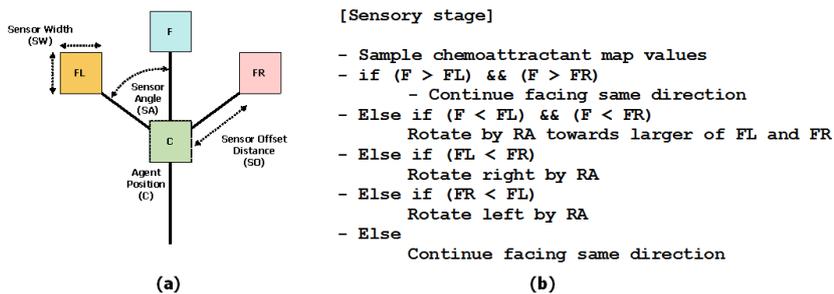

Figure 9. Architecture of a single component of the virtual plasmodium and its sensory algorithm. (a) Morphology showing agent position 'C' and offset sensor positions (FL, F, FR), (b) Algorithm for particle sensory stage.

### 6.2. Problem Representation

Path start and end points were represented by projection of chemo-attractant to the diffusive lattice at their positions indicated by white pixels on the source configuration images. Lattice size varied with each particular arena but varied from 150 pixels minimum size to 416 pixels maximum size. The attractant projection concentration was 6.375 units per scheduler step. Projection of repellent sources from arena boundaries (used in collision-free planning experiments) was implemented by negatively valued projection ($-6.375$ units) into the lattice at arena boundary locations causing repulsion of the blob from these regions. Uncovering of obstacles by the shrinking blob acted to project repellent into the lattice ($-6.375$ units) at exposed areas of obstacles, causing the blob to be repelled from these regions. Each pixel comprising the obstacle was deemed to be uncovered if there were any agent particles within an $11 \times 11$ window of the pixel. If an obstacle was covered by particles the repellent projection was suppressed by reducing the projection of repellent to $-0.006375$ units. This ensured that repulsion from obstacles only occurred when a significant part of the obstacle had been uncovered by the shrinking blob. Diffusion in the lattice was implemented at each scheduler step and at every site in the lattice via a simple mean filter of kernel size $3 \times 3$. Damping of the diffusion distance, which limits the distance of chemo-attractant gradient diffusion, was achieved by multiplying the mean kernel value by 0.9 per scheduler step.

The blob was initialised by creating a population of particles and inoculating the population within the habitable regions of the arena The exact population size differed depending on the size of the arena (and habitable area) initial blob was typically composed of between 24000 and 70000 particles. Particles were given random initial positions within the habitable area and also random initial orientations. Particle sensor offset distance (SO) was 7 pixels. Angle of rotation (RA) was set to $45°$ and sensor angle (SA) was set to $90°$. Agent forward displacement was 1 pixel per scheduler step and particles moving forwards successfully deposited 5 units of chemo-attractant into the diffusive lattice. This value is slightly less than the attractant projection value, causing the particles to be anchored to projection sites and ultimately constraining the shape of the shrinking blob. Both data projection stimuli and agent particle trails were represented by the same chemo-attractant ensuring that the particles were attracted to both data stimuli and other agents' trails. The collective behaviour of the particle population was cohesion, minimisation, and morphological adaptation to the configuration of stimuli.

### 6.3. *Shrinkage Mechanism*

Adaptation of the blob size was implemented via tests at regular intervals. Growth of the population was implemented as follows: If there were between 1 and 10 particles in a $9 \times 9$ neighbourhood of a particle, and the particle had moved forward successfully, the particle attempted to divide into two if there was a space available at a randomly selected empty location in the immediate $3 \times 3$ neighbourhood surrounding the particle. Shrinkage of the population was implemented as follows: If there were between 0 to 79 particles in a $9 \times 9$ neighbourhood of a particle the particle survived, otherwise it was deleted. Deletion of a particle left a vacant space at this location which was filled by nearby particles (due to the emergent cohesion effects), thus causing the blob to shrink slightly. As the process continued the blob shrunk further and adapted its shape to the stimuli provided by the configuration of path source points, arena boundaries and repellent obstacles. The frequency at which the growth and shrinkage of the population was executed determined the turnover rate for the population. The frequency of testing for particle division was every 10 scheduler steps and the frequency for testing for particle removal was every 2 scheduler steps. Since the shrinking blob method is only concerned with the reduction in size of the population it might be asked as to why there were tests for particle division at all. The particle division mechanism was present to ensure that the adaptation of the blob was uniform across the sheet to prevent 'tears' or holes forming within the blob sheet, particularly at the start of an experiment before flux within the blob was initially stabilised.